\documentclass[%
 reprint,
superscriptaddress,
 amsmath,amssymb,
 aps,
]{revtex4-2}

\usepackage{xcolor}
\usepackage{graphicx}
\usepackage{dcolumn}
\usepackage{bm}
\usepackage{hyperref}

\definecolor{addblue}{rgb}{0.00,0.20,0.75}

\begin{document}
\raggedbottom
\preprint{APS/123-QED}


\title{
Gamma-Ray Bursts from Primordial Black Hole-Induced Neutron Star Collapse}


\author{Yu Watanabe}
\email{yuwatanabe@g.ucla.edu}
\affiliation{Department of Physics and Astronomy, University of California, Los Angeles \\ Los Angeles, California, 90095-1547, USA}
\author{Koichiro Yasuda}
\email{yasuda@ucla.edu}
\affiliation{Department of Physics and Astronomy, University of California, Los Angeles \\ Los Angeles, California, 90095-1547, USA}
\author{Alexander Kusenko}
\email{kusenko@ucla.edu}
\affiliation{Department of Physics and Astronomy, University of California, Los Angeles \\ Los Angeles, California, 90095-1547, USA}
\affiliation{Kavli Institute for the Physics and Mathematics of the Universe (WPI), UTIAS \\The University of Tokyo, Kashiwa, Chiba 277-8583, Japan}

\begin{abstract}
\noindent
We show that the capture and subsequent endoparasitic consumption of a neutron star (NS) by a primordial black hole (PBH) can produce a short gamma-ray burst (GRB) without forming an accretion disk. During PBH-induced NS consumption, the NS spin is amplified through angular-momentum conservation and transfer from infalling matter, while flux freezing and a differential-rotation-driven dynamo amplify the magnetic field. We find that, for rapidly rotating millisecond pulsars (MSPs), these mechanisms can power short GRBs through magnetic dipole radiation and magnetic energy release. If asteroid-mass PBHs in the range $10^{-14}M_\odot$--$10^{-11}M_\odot$ constitute the dark matter, we estimate that searches of the Fermi-GBM archive and a future MeV gamma-ray mission such as COSI could detect these bursts. We also discuss possible multimessenger counterparts, including fast radio bursts powered by magnetic energy release, kilonovae and $511\,{\rm keV}$ line emission associated with centrifugally ejected neutron-rich matter, and gravitational waves from the rapid collapse and subsequent black-hole ringdown. Together, these signals probe complementary stages of PBH-induced NS consumption.
\end{abstract}


\maketitle

\section{Introduction}

Uncovering the nature of dark matter (DM) remains one of the central challenges in astrophysics, cosmology, and particle physics. Primordial black holes (PBHs) are compelling DM candidates. 
Although observations constrain PBH abundances over a broad range of masses~\cite{Carr:2009jm,Niikura:2017zjd,Niikura:2019kqi,Zumalacarregui:2017qqd,Boudaud:2018hqb,Quinn:2009zg,Carr:2018rid,Poulin:2017bwe,Serpico:2020ehh,Inoue:2017csr,Lu:2020bmd,Raidal:2017mfl}, the asteroid-mass window of $10^{17}$--$10^{23}\,{\rm g}$ remains  unconstrained~\cite{Carr:2020gox,Carr:2021bzv}.  Neutron stars (NSs) offer a promising way to probe this window~\cite{Kouvaris:2013kra,Capela:2013yf,Fuller:2017uyd}. In dense environments, PBHs can be captured by NSs, consume them from within, and induce their collapse, potentially producing multiple observable signatures, including fast radio bursts (FRBs)~\cite{Fuller:2017uyd,Abramowicz:2017zbp,Kainulainen:2021rbg}, kilonovae, and $511\,{\rm keV}$ line emission~\cite{Fuller:2017uyd}. In this Letter, we show that such a collapse can also produce a short gamma-ray burst (sGRB) without an accretion disk and identify its distinctive observational signatures.

\section{PBH capture by NS}

A PBH can be captured by an NS if it loses sufficient kinetic energy through dynamical friction and accretion during the encounter. The capture rate per NS is given by $F=f_{\rm PBH}F_0$, where $f_{\rm PBH}$ is the PBH fraction of DM density $\rho_{\rm DM}$ and, and $F_0$ is~\cite{Caiozzo:2024flz,Genolini:2020ejw,Capela:2013yf}
\begin{equation}
    F_0 = \sqrt{6\pi} \frac{\rho_{\rm DM}}{M_{\rm PBH}}
    \left[ \frac{R_{\rm NS} R_{\rm S} c^2}{v (1-R_{\rm S}/R_{\rm NS})} \right]
    \left(1-e^{-E_{\rm loss}/E_b}\right),
\end{equation}
where $M_{\rm PBH}$ and $v$ are the PBH mass and velocity, respectively, and $R_{\rm S}=2GM_{\rm NS}/c^2$ is the Schwarzschild radius of the NS, with $c$ being the speed of light. Throughout this work, we adopt a representative NS radius $R_{\rm NS}=12\,{\rm km}$ and mass $M_{\rm NS}=1.6\,M_\odot$~\cite{Shapiro:1983du}. The energy loss during a single transit is estimated as $E_{\rm loss} \simeq 67 \, G^2M_{\rm PBH}^2M_{\rm NS}/(R_{\rm NS}^2 c^2)$, while the characteristic PBH kinetic energy is $E_b=M_{\rm PBH}v^2/3$. 

After capture, the PBH settles into a fully contained orbit around the NS center on a timescale of $\simeq 3.8 \times 10^4 (10^{22}\,{\rm g}/M_{\rm PBH})^{3/2}\,{\rm yr}$~\cite{Caiozzo:2024flz}. The PBH subsequently consumes the NS through spherical Bondi accretion~\footnote{NS rotation does not impede this quasi-spherical accretion~\cite{Kouvaris:2013kra}: The specific angular momentum at the innermost stable orbit (ISCO) is $j_{\rm ISCO}=2\sqrt{3}\psi GM_{\rm PBH}/c$, where $\psi=1$ ($1/3$) for a Schwarzschild (extreme Kerr) BH. The Bondi radius is $r_B = GM_{\rm PBH}/v_s^2$, so the specific angular momentum of the inflowing matter satisfies $\Omega r_B^2/j_{\rm ISCO} \sim 0.07 (\Omega/10^4 s^{-1})(M_{\rm PBH}/M_{\rm NS})(1/\psi) \ll 1$}. The accretion rate is given by $dM_{\rm PBH}(t)/dt = 4\pi\lambda\, G^2M_{\rm PBH}(t)^2\rho_c/v_s^3 \equiv kM_{\rm PBH}(t)^2$. Hereafter, a time argument is included when necessary to explicitly indicate the time evolution of a quantity. We model the NS with an AP3 equation of state~\cite{Akmal:1998cf}, which is consistent~\cite{Biswas:2024hja} with current observations of NS masses, radii, and tidal deformabilities~\cite{Demorest:2010bx,Antoniadis:2013pzd,Miller:2021qha,Riley:2021pdl,LIGOScientific:2017vwq}. We adopt a typical central density $\rho_c=8.4\times 10^{14}\,{\rm g\,cm^{-3}}$~\cite{Schnauck:2021hlm} and the accretion eigenvalue $\lambda\simeq1.7$ of relativistic Bondi flow~\cite{Schnauck:2021hlm, Richards:2021zbr} together with the central sound speed $v_s\simeq0.75\,c$~\cite{Read:2008iy, Schnauck:2021hlm}. Solving the Bondi equation yields $M_{\rm PBH}(t) = M_{\rm PBH}(0)/[1-k \, M_{\rm PBH}(0) \, t]$, from which the NS consumption time follows as $t_{\rm con} \simeq 0.5 \, (10^{22}\,{\rm g}/M_{\rm PBH})\,{\rm yr}$. Once the BH mass reaches a fraction of the NS mass, $\epsilon M_{\rm NS}$, the remaining accretion time becomes $\Delta t \simeq 0.4 \, (0.1/\epsilon)\,{\rm ms}$. This is comparable to the stellar dynamical time, $t_{\rm dyn}\sim\sqrt{R_{\rm NS}^3/(GM_{\rm NS})} \simeq 0.1\,{\rm ms}$, implying that most of the stellar mass is consumed during a millisecond-scale endgame, effectively in free fall.

\section{Dynamics during NS Consumption: Spin and Magnetic-Field Amplification}\label{sec:dynamics}

As the PBH consumes the NS, the inner layers contract faster than the outer layers, generating differential rotation through angular momentum conservation. However, viscosity~\cite{Markovic:1994bu,Kouvaris:2013kra} and magnetic stresses~\cite{Markovic:1994bu} transport angular momentum outward. The kinematic viscosity of a typical NS is $\nu=2\times10^{11}(T/10^5\,{\rm K})^{-2}\,{\rm cm^2\,s^{-1}}$, where $T$ is the NS temperature~\cite{1979ApJ...230..847F, 1987ApJ...314..234C, van1991neutron}, giving an angular momentum transport timescale of $t_\nu\sim R_{\rm NS}^2/\nu\sim7(T/10^5\,{\rm K})^2\,{\rm s}$. Magnetic stresses act on the Alfv\'en timescale $t_A\sim R_{\rm NS}/v_A\sim0.1(10^{15}\,{\rm G}/B_p)\,{\rm s}$, where $v_A=B_p/\sqrt{4\pi\rho_c}$ is the Alfv\'en speed and $B_p$ is the poloidal magnetic field~\cite{Shapiro:2000zh,Cook:2003ku,Duez:2004nf}. On timescales longer than these transport timescales, the NS maintains nearly rigid-body rotation. During the final stage, when $\Delta t<\min(t_\nu,t_A)$, differential rotation develops within the region where the BH gravity exceeds the self-gravity of the enclosed NS matter, namely $r<r_{\rm DR}=5(\epsilon/0.1)^{1/3}\,{\rm km}$.

The NS spin is amplified through angular momentum conservation during contraction and by angular momentum transfer from the infalling matter. In the rigid-body rotation regime, the angular momentum is $J_{\rm NS}=I\Omega$, where $I\propto M_{\rm NS}R_{\rm NS}^2$ is the moment of inertia and $\Omega$ is the angular frequency. Efficient angular momentum transport keeps the BH spin contribution negligible~\cite{Kouvaris:2013kra}, implying $\Omega(t)\propto[M_{\rm NS}(t)R_{\rm NS}^2(t)]^{-1}$. Since both $M_{\rm NS}$ and $R_{\rm NS}$ remain nearly constant until the ms-scale endgame~\cite{Lattimer:2000nx,Kruger:2023olj,Kruger:2025sda}, the spin-up remains modest because $t_A$ ($t_\nu$) is much longer than the endgame unless $B_p$ ($T$) is exceptionally large (small). Once differential rotation develops, contraction dominates the spin evolution, yielding $\Omega(t)\simeq\Omega(0)[R_{\rm NS}(0)/R_{\rm NS}(t)]^2$. During the final stage, $R_{\rm NS}$ approaches the outer Kerr horizon, $R_+$, of the BH with $M_{\rm BH}\sim M_{\rm NS}(0)$. Depending on the dimensionless BH spin parameter $a\simeq cJ_{\rm NS}/(GM_{\rm NS}^2)$, $R_+\simeq(0.5$--$1)R_{\rm S}$, implying an amplification of $\Omega$ by a factor of $\sim6$--$25$.

Differential rotation amplifies the magnetic field through an $\alpha\Omega$ dynamo~\cite{Brandenburg:2004jv,Reboul-Salze:2021rmf,Kiuchi:2023obe}. Shear first winds the poloidal field into a toroidal component ($\Omega$ effect), giving $B_\phi(t)\sim\Omega \, t \, B_p$~\cite{Brandenburg:2004jv,Duez:2006qe}. The resulting magnetized, differentially rotating flow is unstable to the magnetorotational instability (MRI), which exponentially amplifies magnetic perturbations and drives turbulence at a timescale of rotaion~\cite{Balbus:1991ay,Balbus:1998ja}. MRI-driven turbulence subsequently regenerates a large-scale poloidal field through the $\alpha$ effect on a timescale of $\mathcal{O}(10)\,{\rm ms}$~\cite{Kiuchi:2023obe}. Efficient dynamo amplification begins once $r_{\rm DR}\gtrsim\lambda_{\rm MRI}$, where $\lambda_{\rm MRI}$ is the wavelength of the fastest-growing MRI mode, corresponding to $\Delta t_{\rm MRI}\sim 200\,(10^{15}\,{\rm G}/B)^3 (\Omega/10^4\,{\rm s^{-1}})^3\,{\rm s}$. The available time for dynamo amplification is therefore $\Delta t_{\alpha\Omega}=\min(\Delta t_{\rm MRI},t_\nu,t_A)$. NSs are broadly classified into slowly rotating systems, such as ordinary radio pulsars and magnetars with $\Omega=\mathcal{O}(1)\,{\rm s^{-1}}$, and rapidly rotating millisecond pulsars (MSPs) with $\Omega=\mathcal{O}(10^3)\,{\rm s^{-1}}$~\cite{Manchester:2004bp}. While the former have insufficient time for magnetic winding and MRI growth, MSPs are expected to undergo efficient dynamo amplification since $\Omega\Delta t_{\alpha\Omega}\gg1$.

The energy reservoir for the dynamo is the differential rotation energy of the NS, $E_{\rm diff}$. Mass and angular-momentum conservation give the rotation profile $\Omega(r)=\Omega(0)(1+r_{\rm DR}^3/r^3)^{2/3}$, leading to $E_{\rm diff}\sim4\rho\,\Omega(0)^2r_{\rm DR}^5$, most of which is stored within $r_{\rm DR}$. Recent simulations find that a fraction $\epsilon_B=\mathcal{O}(10^{-2}\text{--}10^{-1})$ of $E_{\rm diff}$ is converted into magnetic energy~\cite{Shapiro:2000zh,Reboul-Salze:2020mnw,Kiuchi:2023obe}, while the poloidal component contains a fraction $\epsilon_p=\mathcal{O}(10^{-2}\text{--}10^{-1})$ of the total magnetic energy~\cite{Reboul-Salze:2020mnw,Reboul-Salze:2021rmf}. We adopt $\epsilon_B\sim10^{-1}$ and $\epsilon_p\sim10^{-1}$. The dynamo operates in the stellar interior, but we assume that buoyancy causes the amplified flux tubes to rise at $v_A\sim1.4\sqrt{\epsilon_B}\,r_{\rm DR}\Omega(0)$~\cite{1955ApJ...121..491P,1975ApJ...198..205P}. Stratification may slow this rise~\cite{Reisenegger:2008yk}. The requirement $R_{\rm NS}/v_A\lesssim\Delta t$ determines the latest time at which the amplified magnetic field can begin rising and reach the surface before collapse. Equating the dipole energy, $B_p^2R_{\rm NS}^3/4$, to $\epsilon_B\epsilon_p E_{\rm diff}$ then yields a surface poloidal field at collapse of $B_p\sim10^{16}(\Omega(0)/10^4\,{\rm s^{-1}})^{9/4}\,{\rm G}$. Magnetic flux freezing further amplifies the surface poloidal field according to $B_p(t)=B_p\,[R_{\rm NS}(0)/R_{\rm NS}(t)]^2$.

\section{GRB}\label{sec:GRB}

Recently, it was proposed that PBH--NS systems can produce short GRBs if an accretion disk forms during the final stage of NS consumption~\cite{Takhistov:2017nmt}, with jets powered by disk-assisted mechanisms~\cite{Eichler:1989ve,blandford1982hydromagnetic,blandford1977electromagnetic}. However, disk formation remains uncertain~\cite{Margalit:2015qza,East:2019dxt}, and even if a disk forms, these mechanisms may not provide sufficient jet energy~\cite{Takhistov:2017nmt}. A necessary condition for disk formation is that matter remains outside the innermost stable circular orbit, $R_{\rm ISCO}$, of the resulting BH. We first consider rigid rotation. For a rapidly rotating NS near the mass-shedding limit, $\Omega\simeq\Omega_{\rm K}\equiv\sqrt{GM_{\rm NS}/R_{\rm NS}^3}\simeq1.1\times10^4\,{\rm s^{-1}}$, the equatorial radius exceeds the polar radius, with $R_{\rm eq}=(3/2)R_p$~\cite{Shapiro:2004zx}. After the BH has consumed the NS down to $R_p$, the remaining equatorial radius is $R_{\rm eq}^f=R_{\rm eq}-R_p+R_+$. Requiring $R_{\rm eq}^f>R_{\rm ISCO}$ for disk formation~\cite{Shapiro:1983du,Shapiro:2004zx} yields $\Omega(0)>7.1\times10^3\,{\rm s^{-1}}$. By contrast, the fastest known pulsar, PSR~J1748$-$2446ad, has an angular velocity of only $\Omega(0)\simeq4.5\times10^3\,{\rm s^{-1}}$~\cite{Hessels:2006ze}. Differential rotation further suppresses disk formation by limiting outward angular momentum transport, consistent with numerical simulations showing nearly complete NS consumption~\cite{East:2019dxt}.

We therefore consider whether an NS--PBH system can produce a GRB without forming an accretion disk. PBH consumption amplifies both $\Omega$ and $B_p$, enabling magnetic dipole radiation with luminosity~\cite{Spitkovsky:2006np},
\begin{align}
    & L_{\rm md} \simeq \frac{B_p^2 R_{\rm NS}^6 \Omega^4}{4c^3} (1+\sin^2\alpha) \nonumber \\
    & \simeq 4\times 10^{52} \,{\rm erg/s} \left(\frac{B_p}{10^{16}\,{\rm G}} \right)^{\!\!2}\!
    \left(\frac{R_{\rm NS}}{12\,{\rm km}} \right)^{\!\!6}\!
    \left(\frac{\Omega}{10^4\,{\rm s^{-1}}} \right)^{\!\!4},
    \label{eq:Emd}
\end{align}
where we adopt an inclination angle $\alpha\simeq\pi/4$. With the dynamo-amplified field, the luminosity just before collapse is $L_{\rm md}\sim4\times10^{52}(\Omega(0)/10^4\,{\rm s^{-1}})^{17/2}\,{\rm erg/s}$. During the final contraction, angular momentum conservation and magnetic flux freezing imply $L_{\rm md}\propto R_{\rm NS}^{-6}$, enhancing the luminosity by a factor of $[R_{\rm NS}(0)/R_+]^6\sim3\times10^2$--$2\times10^4$. This enhancement allows even the fastest known pulsar to release GRB-scale energy, $E_{\rm GRB}\sim10^{48}\,{\rm erg}$, on a submillisecond timescale, making MSP--PBH systems promising GRB progenitors. The emission may be further enhanced by the opening of field lines through twists induced by the toroidal field~\cite{Parfrey:2013gza}, by additional poloidal-field generation through Babcock--Leighton-like processes~\cite{1961ApJ...133..572B,1969ApJ...156....1L}, or by prolonged dynamo activity if accretion is delayed~\cite{Kouvaris:2013kra,Cho:2024wsp}. By contrast, slowly rotating NSs produce much weaker emission because of their lower $\Omega$ and less efficient dynamo amplification.

The magnetic-dipole outflow may thermalize at a pair photosphere, producing a quasi-Planck spectrum peaking at $\sim1\,{\rm MeV}$ with a nonthermal tail extending to $\sim10^2\,{\rm MeV}$~\cite{Usov:1992zd}. The endgame dynamics determine the light curve: for near-free-fall contraction, $R_{\rm NS}\propto \Delta t^{2/3}$, giving $L_{\rm md}\propto \Delta t^{-4}$. The pulse is then terminated by horizon formation on a magnetospheric ``balding'' timescale of tens of $GM_{\rm BH}/c^3\sim\mathcal{O}(10)\mu{\rm s}$, followed by ringdown of the collapsing magnetosphere~\cite{Falcke:2013xpa,Most:2024qgc}. A millisecond-scale MeV transient with a steep power-law rise and an abrupt $\mathcal{O}(0.1)\,{\rm ms}$ cutoff is thus a characteristic signature of this scenario, distinguishable from both binary-merger short GRBs and extragalactic magnetar giant flares, whose light curves instead rise on $\mu$s timescales and subsequently decay~\cite{Roberts:2021gfl,Mereghetti:2024m82}.

This engine resembles that proposed by Usov~\cite{Usov:1992zd,usov1994nature}, who considered the accretion-induced collapse of a strongly magnetized white dwarf. Such a scenario requires an O--Ne progenitor to avoid a thermonuclear explosion~\cite{Nomoto:1991zz}, sufficiently large $B_p$ and $\Omega$, and faces baryon loading from neutrino-driven winds of the hot proto-NS~\cite{Dessart:2006zz}. PBH consumption alleviates these difficulties: an old, cold MSP undergoes spin and magnetic-field amplification, while nearly complete NS consumption may suppress baryon loading. Even if magnetic dipole emission is inefficient, horizon formation can release energy stored in the exterior magnetosphere~\cite{Lehner:2011aa,Most:2024qgc}, potentially providing an additional GRB energy of $E_{\rm GRB}\sim10^{48}(\epsilon_M/0.1)(B_p/10^{16}\,{\rm G})^2\,{\rm erg}$, where $\epsilon_M$ is the fraction of the exterior magnetospheric energy converted into gamma rays. Moreover, buoyant toroidal magnetic flux emerging through the stellar surface before collapse can release energy through reconnection~\cite{Kluzniak:1997nt,Ruderman:2000np}. Unlike dipole radiation, this channel can tap the total magnetic energy $\epsilon_BE_{\rm diff}$, yielding $E_{\rm GRB}\sim3\times10^{49}(\eta_T/0.1)(\Omega(0)/10^4\,{\rm s^{-1}})^{9/2}\,{\rm erg}$, where $\eta_T$ is the fraction of the emerged toroidal-field energy radiated in gamma rays.

\begin{figure}[t]
    \centering
    \includegraphics[keepaspectratio, scale=0.72]{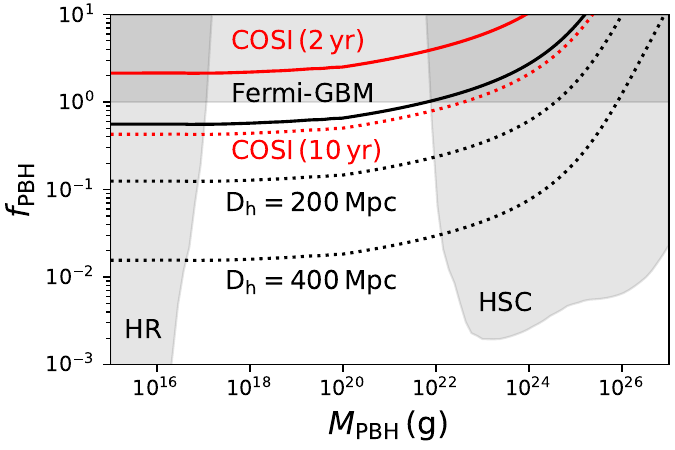}
    \caption{Parameter space in which at least one GRB from MSP--PBH systems is expected. Dotted curves show the thresholds for different $D_{\rm h}$, assuming $f_{\rm sky}=1$ and $t_{\rm obs}=2\,{\rm yr}$. The black solid curve shows the detection threshold for Fermi-GBM, while the red solid and dotted curves correspond to COSI observations of 2 and 10 years, respectively. Shaded regions indicate $f_{\rm PBH}>1$ and existing constraints from Hawking radiation (HR)~\cite{Carr:2009jm,Laha:2020ivk,Coogan:2020tuf} and Subaru HSC microlensing (HSC)~\cite{Niikura:2017zjd}.}
    \label{fig: fPBH-GRB}
\end{figure}

The expected number of GRBs is
\begin{equation}
    N_{\rm GRB}=f_{\rm PBH}\,F_0\,t_{\rm obs}\,N_{\rm MSP}\,N_{\rm G}\,f_{\rm sky},
    \label{eq:NGRB}
\end{equation}
where $t_{\rm obs}$ is the observation time and $f_{\rm sky}$ is the instantaneous fraction of the sky covered by the instrument. We evaluate $F_0$ over $50\,{\rm GeV/cm^3}\lesssim\rho_{\rm DM}\lesssim8.8\times10^2\,{\rm GeV/cm^3}$~\cite{Burkert:1995yz,Fuller:2017uyd} and $50\,{\rm km/s}\lesssim v\lesssim105\,{\rm km/s}$~\cite{Read:2008fh,Read:2009iv,Kaplinghat:2013xca}, adopting an optimistic benchmark~\cite{Fuller:2017uyd,Takhistov:2017bpt}. A systematic exploration of these astrophysical parameters is left for future work. We account for natal pulsar kicks following Ref.~\cite{Fuller:2017uyd}, assuming a pulsar velocity dispersion of $48\,{\rm km/s}$~\cite{Cordes:1997my}. We adopt the 95\% CL upper limit of $N_{\rm MSP}\sim2\times10^5$ MSPs per Milky Way-like galaxy~\cite{Gautam:2021wqn}. The number of such galaxies within a horizon distance $D_{\rm h}$ is estimated from the blue-light luminosity density as $N_{\rm G}\simeq1.16\times10^{-2}(4\pi/3)(D_{\rm h}/{\rm Mpc})^3$~\cite{Kopparapu:2007ib}. Figure~\ref{fig: fPBH-GRB} shows the parameter space where at least one GRB is expected as a function of $D_{\rm h}$, assuming $f_{\rm sky}=1$ and $t_{\rm obs}=2\,{\rm yr}$, together with existing PBH constraints~\cite{Carr:2009jm,Laha:2020ivk,Coogan:2020tuf,Niikura:2017zjd}.

Fermi-GBM has operated for $t_{\rm obs}\simeq17\,{\rm yr}$ with $f_{\rm sky}\sim0.7$~\cite{vonKienlin:2020xvz}. Guided by the $64\,{\rm ms}$ detection sensitivity reported in Ref.~\cite{Howell:2024wwr}, we assume an effective fluence threshold of $S_{\rm GBM}\sim10^{-7}\,{\rm erg\,cm^{-2}}$. The corresponding luminosity distance is $D_L=\sqrt{E_{\rm GRB}/(4\pi S_{\rm GBM})}$, and we account for cosmological volume and time-dilation effects in calculating the expected event rate~\cite{Hogg:1999ad,Bloom:2001ts}. NASA's upcoming MeV $\gamma$-ray telescope COSI~\cite{Tomsick:2021wed,Tomsick:2023aue} has a nominal observation period of $t_{\rm obs}=2\,{\rm yr}$, with a possible extension. Its germanium detectors cover $f_{\rm sky}\simeq0.25$, and we assume a fluence threshold of $S_{\rm COSI}\sim5\times10^{-8}\,{\rm erg\,cm^{-2}}$, corresponding to at least one detected photon for an effective area of $\sim30\,{\rm cm^2}$~\cite{Gulick:2024clf}. Its BGO shields have a similar sensitivity and an effective sky coverage of $f_{\rm sky}\simeq0.6$~\cite{Neights:2026mdl}. Fig.~\ref{fig: fPBH-GRB} shows the resulting detection thresholds, accounting for the $\Omega$ dependence of $E_{\rm GRB}$ by averaging over a Weibull distribution of MSP spin frequencies~\cite{Patruno:2017oum,Liu:2023qnf}. Fermi-GBM could thus probe part of the currently unconstrained PBH parameter space through PBH-induced GRBs, although the predicted event rates remain subject to substantial astrophysical uncertainties. COSI is less sensitive during its nominal mission but could surpass Fermi-GBM with a 10-year extended mission. Since the expected number of detections scales approximately as $E_{\rm GRB}^{3/2}$, the energy enhancement and additional emission channels discussed above could further improve detectability. Moreover, COSI's $0.2$--$5\,{\rm MeV}$ band encompasses the predicted spectral peak, while its Compton imaging and polarization capabilities~\cite{Kierans:2022eid} could help identify these events and distinguish them from conventional GRBs.

Our predicted GRBs have a spectral peak near $1\,{\rm MeV}$ and an isotropic-equivalent energy of $E_{\rm GRB}\lesssim10^{48}\,{\rm erg}$. At this energy, their spectral peaks are substantially higher than those expected from the type~II (Amati) correlation~\cite{Amati:2002ny,Amati:2006kf} and the distinct type~I relation~\cite{Minaev:2019bqz}, which predict only a few to tens of keV. Unlike these GRBs, whose spectral peaks and radiated energies may depend on common jet properties, the spectral peak in our scenario is governed primarily by pair-photosphere physics, while the total energy is determined by magnetic dipole emission. These hard, millisecond-scale bursts also contrast with the soft X-ray transients observed by \textit{Einstein Probe} and SVOM~\cite{Yuan:2022bfl,Wei:2016eox}, some of which are associated with type~Ic supernovae~\cite{Campana:2006qe,Sun:2024ozr,Srinivasaragavan:2025wcn}. The absence of a stellar envelope and the nearly complete consumption of the NS favor a baryon-poor outflow, making strong spectral softening less likely. Furthermore, the abrupt termination at BH formation distinguishes this scenario from transients with extended emission or X-ray plateaus~\cite{Rowlinson:2013ue,Zhang:2024epc,OConnor:2025epc}.

\section{Other signals}\label{sec:other}

MSP--PBH systems may also produce counterparts to GRBs, providing complementary probes of the underlying physical processes. One possible counterpart is an FRB~\cite{Fuller:2017uyd,Abramowicz:2017zbp}. The no-hair theorem implies that the magnetic field initially supported by the NS cannot remain anchored to the final BH, releasing part of the magnetospheric energy during collapse~\cite{Israel:1967za,Carter:1971zc}. For sufficiently strong magnetic fields and rapid rotation, copious pair creation makes the inner magnetosphere opaque, suppressing radio emission~\cite{Sturrock:1971zc,Schwinger:1951nm,Nishiura:2026gsj}. An external relativistic shock therefore provides a plausible FRB emission site~\cite{Falcke:2013xpa,Beloborodov:2022pvn,Most:2024qgc}. Assuming that $\sim10^{-2}$ of the magnetic energy $E_M\sim10^{49}(B_p/10^{16}\,{\rm G})^2\,{\rm erg}$ powers a relativistic shell, synchrotron maser emission with efficiency $\sim10^{-3}$~\cite{Plotnikov:2019zuz,Sironi:2021wca} yields $E_{\rm FRB}\sim10^{44}(B_p/10^{16}\,{\rm G})^2\,{\rm erg}$, consistent with observed FRBs for moderate $B_p$~\cite{Petroff:2019tty}. The characteristic frequency and duration are $\nu\simeq3\Gamma\nu_p$ and $t_{\rm FRB}\simeq R_{\rm dec}/(2c\Gamma^2)$, where $\Gamma$ is the shock Lorentz factor, $\nu_p\simeq9\times10^3\sqrt{n_{\rm ext}/{\rm cm^{-3}}}\,{\rm Hz}$ is the upstream plasma frequency, and $R_{\rm dec}$ is the deceleration radius~\cite{Plotnikov:2019zuz,Meszaros:2006rc}. For a central molecular zone (CMZ)-like environment with $n_{\rm ext}\sim10^4\,{\rm cm^{-3}}$, $\Gamma\sim700$ yields GHz emission, with millisecond duration for $R_{\rm dec}\sim3\times10^{13}\,{\rm cm}$. The neutron-rich ejecta discussed below could absorb GHz radiation if distributed. However, preferential consumption of polar matter and equatorial concentration of ejecta may create a baryon-clean polar funnel, allowing both GRB and FRB emission to escape along polar directions. No confirmed GRB--FRB association has yet been observed~\cite{Cunningham:2019xbu}; detecting such an association would provide compelling evidence for the MSP--PBH scenario.

Although most of the NS is consumed by the PBH, efficient angular momentum transport may allow rapidly rotating MSPs to eject up to $\mathcal{O}(10^{-1})\,M_\odot$ of neutron-rich matter~\cite{Fuller:2017uyd}. Since such transport becomes inefficient during the endgame (Sec.~\ref{sec:dynamics}), we regard this as an upper limit and adopt a fiducial ejecta mass of $\mathcal{O}(10^{-2})\,M_\odot$. Decompression of the neutron-rich ejecta triggers $r$-process nucleosynthesis~\cite{Lattimer:1977igd,meyer1989decompression,Fuller:2017uyd}, followed by radioactive heating that produces kilonova-like emission~\cite{Li:1998bw,Metzger:2010sy,Hotokezaka:2015eja}. Heating to $\sim0.1\,{\rm MeV}$ may also produce thermal electron--positron pairs~\cite{Goriely:2011er}. For an NS disruption rate of $10^{-5}\,{\rm yr^{-1}}$ in the Galactic Center, the positron production rate is $R_{e^+}\sim10^{50}(M_{\rm ej}/0.1\,M_\odot)\,{\rm yr^{-1}}$~\cite{Fuller:2017uyd}, potentially contributing to the observed $511\,{\rm keV}$ emission~\cite{Siegert:2015knp,Kierans:2019aqz} without violating stringent in-flight annihilation constraints~\cite{Beacom:2005qv}.

The millisecond endgame may also generate GWs through deviations from spherical symmetry, including NS rotation and off-center PBH accretion, with characteristic frequencies $f_{\rm GW}\sim\mathcal{O}(1)\,{\rm kHz}$~\cite{Baiotti:2007np,Horowitz:2019aim,Baumgarte:2024mei}. The final BH produces ringdown GWs at $\sim8$--$9\,{\rm kHz}$~\cite{East:2019dxt}. These frequencies lie near or above the most sensitive bands of current detectors such as LIGO and KAGRA~\cite{LIGOScientific:2014pky,KAGRA:2025yfg}, but may be probed by future detectors including CE, ET, and NEMO~\cite{Reitze:2019iox,Punturo:2010zz,Hild:2010id,Ackley:2020atn}. Assuming a GW emission efficiency of $\sim10^{-6}$ suggested by numerical simulations~\cite{Baiotti:2007np}, the radiated energy is $E_{\rm GW}\sim3\times10^{48}\,{\rm erg}$, corresponding to $h_{\rm rss}\sim10^{-21}(1\,{\rm kHz}/f_{\rm GW})(10\,{\rm kpc}/d)\,{\rm Hz^{-1/2}}$~\cite{Sutton:2013ooa}. Thus, Galactic MSP--PBH events may be detectable by future GW observatories.

\section{Conclusion}

We have shown that PBH-induced NS consumption can produce short GRBs without an accretion disk. PBH-induced contraction amplifies the stellar spin and magnetic field, potentially powering GRBs through magnetic dipole radiation and magnetic energy release, particularly in MSPs. If asteroid-mass PBHs constitute $\mathcal{O}(0.1)$ of DM, searches of the Fermi-GBM archive and an extended 10-year COSI mission could detect one or more such events under optimistic assumptions. Possible multimessenger counterparts, including FRBs, kilonovae, $511\,{\rm keV}$ emission, and GWs, provide complementary probes of the asteroid-mass PBH DM window.

\begin{acknowledgments}

Y.W. was supported by the JSPS Overseas Research Fellowships; and the U.S.\ Department of Energy under Award No.\ DE-SC0009937.
K.Y. was supported by the LANL-SoCal Hub Graduate Fellowship program and the UC SoCal Hub.
A.K.\ was supported by the U.S.\ Department of Energy (DOE) under Grant No.\ DE-SC0009937 and by the World Premier International Research Center Initiative (WPI), MEXT, Japan (Kavli IPMU).

\end{acknowledgments}

\bibliography{refs}

\end{document}